# ICEYE MICROSATELLITE SAR CONSTELLATION STATUS UPDATE: EVALUATION OF FIRST COMMERCIAL IMAGING MODES

*Vladimir Ignatenko, Pekka Laurila, Andrea Radius, Leszek Lamentowski, Oleg Antropov, Darren Muff*

ICEYE Oy, Maarintie 6, 02150 Espoo, Finland

## ABSTRACT

The ICEYE constellation features the first operational microsatellite based X-band SAR sensors suitable for allweather day-and-night Earth Observation. In this paper we report on the status of the ICEYE Constellation and describe the characteristics of the first operational imaging modes.

## 1. INTRODUCTION

The ICEYE Constellation is a system intended for persistent near real time monitoring anywhere on Earth. Presently the commercial constellation consists of 3 second-generation ICEYE X-band SAR satellites, providing nearly daily access to any location on the globe. The main benefit of the microsatellite approach is the ability to operate a large number of spacecraft with a reasonable cost. Further plans see the ICEYE Constellation growing with enough satellites to enable persistent access every hour and to operate in multi-static configurations that allow instantaneous coherent collections for elevation modelling or surface velocity measurements.

The first ICEYE satellite, ICEYE-X1, was launched in January 2018. This demonstrator mission produced the world's first SAR imagery with a microsatellite-scale satellite. The first ICEYE SAR instrument used 60MHz of pulse bandwidth, allowing for 5-10 meter resolution imagery. While this resolution was already deemed useful in some applications, such as sea ice monitoring [1], vessel detection or land-use classification [2], the development continued for the next generation 300MHz bandwidth instrument, targeting the full spectrum of commercial SAR applications from 20meter resolution wide scans to sub-meter resolution spotlight imagery.

The first among second-generation ICEYE satellites, ICEYE-X2, was launched 11 months later, in December 2018. With the LEOP (Launch and Early Orbit phase) experience gathered from the previous mission, the team was able to produce the first imagery only 4 days after launch. The commissioning to commercial operations with calibrated imaging modes began after initial commissioning and deliveries to users were started in February 2019. Further two additional satellites were added to the constellation in July 2019, and deliveries of operational imagery to users began in September 2019.

The aim of this paper is to analyze the first commercial imaging modes produced by the ICEYE Constellation utilizing ICEYE-X2 spacecraft and sensor as an example. The sensor parameters and imaging modes are described in Section 2. Image Parameters and Products are further described in Section 3. Typical use-case scenarios for the ICEYE data are summarized in Section 4 and the paper is concluded in Section 5.

## 2. ICEYE SAR SENSOR SPECIFICATIONS AND IMAGING MODES

ICEYE SAR satellites are side-looking X-band SAR sensors utilizing active phased array antenna technology. The sensors and spacecraft are designed to allow Stripmap, Spotlight and ScanSAR imaging modes with the ability to task both left- and right-looking acquisitions. ICEYE-X2, a typical representative instrument of the ICEYE constellation, is currently located on a sun-synchronous polar orbit with 17 days of ground track repeat cycle with 15 imaging orbits per day. All satellites starting from ICEYE-X2 have on-board electric propulsion, enabling all the orbits to be adjusted towards the optimal constellation arrangement as new satellites are launched.

Table 1 lists system and orbital parameters of presently available SAR instruments.

ICEYE satellites collect SAR images in several standard imaging modes. ICEYE's satellites are capable of imaging in Stripmap, Spotlight and ScanSAR modes, with Stripmap and Spotlight imagery operationally available for acquisition worldwide. Notably, all imaging beam modes are available in both right- and left-looking configuration. In Stripmap mode, the ground swath is illuminated with a continuous sequence of pulses while the antenna beam is fixed in elevation and azimuth. This results in an image strip with a continuous image quality in the flight direction. Figure 1 shows an example of stripmap multilooked and ground range detected image. In Spotlight mode, mechanical antenna steering in the azimuth direction is used to increase the illumination time, resulting in an increased synthetic aperture and, therefore, better azimuth resolution compared to a continuous Stripmap mode. Figure 2 shows an example of a multilooked and ground range detected image with resolution of 25 cm. Further, we describe the available imaging modes, and their radiometric and geometric characteristics. A summary of ICEYE imaging modes are gathered in Table 3.

*Table 1. System and orbital parameters of the ICEYE sensors (with ICEYE-X2 as an example)*

| SYSTEM PARAMETER | SPECIFICATION VALUE |
|---|---|
| carrier frequency | 9.65 GHz |
| look side | both LEFT and RIGHT |
| antenna size | 3.2 m x 0.4 m |
| PRF | 2-10 kHz |
| range bw | 40-300 MHz |
| peak power | 4 kW |
| polarization | VV |
| incidence angle (stripmap) | 10-30 |
| incidence angle (spotlight) | 20-35 |
| mass | 85 kg |
| ORBITAL AND ATTITUDE PARAMETERS | SPECIFICATION VALUE |
| nominal orbit height at the equator | 570 km |
| inclination | 97.69 degrees |
| orbits per day | 15 |
| ground track revisit time | 17 days |
| nodal crossing | 10:30 LTDN |

Future developments include the use of various multistatic acquisition modes with synchronized transmit signals [3]. This should enable flexible imaging geometry that can be dynamically adapted to various operational SAR applications.

## 3. ICEYE IMAGE PARAMETERS AND PRODUCTS

In this section, we describe the basic image products and SAR image quality parameters derived from ICEYE SAR data [4].

### 3.1. Standard Products

Standard ICEYE products can be differentiated into two major product groups: Basic Image Products (geo-referenced slant range complex and ground range detected scenes) and Orthorectified Image Products (geo-coded and radiometrically corrected scenes). Basic Georeferenced Image Products are satellite path oriented datasets. They correspond to the Committee on Earth Observation Satellites (CEOS) Level 1b quality and can be ordered either from archive (previously collected imagery) or as planned for future acquisition. Orthorectified Image Products are geo-coded and radiometrically corrected and correspond to CEOS Level 2 quality imagery.

A basic ICEYE product is represented by a set of SAR image binary data along with the corresponding image annotation metadata, delivered as a singular product package. Products are characterized by the payload configuration (such as the imaging mode and look direction) used by the respective satellite, as well as the level of processing that has been applied to the SAR scene. With respect to the data geometric projection and representation, Basic Image Products are differentiated into two primary product types, Single Look Complex (SLC) and Ground Range Detected (GRD).

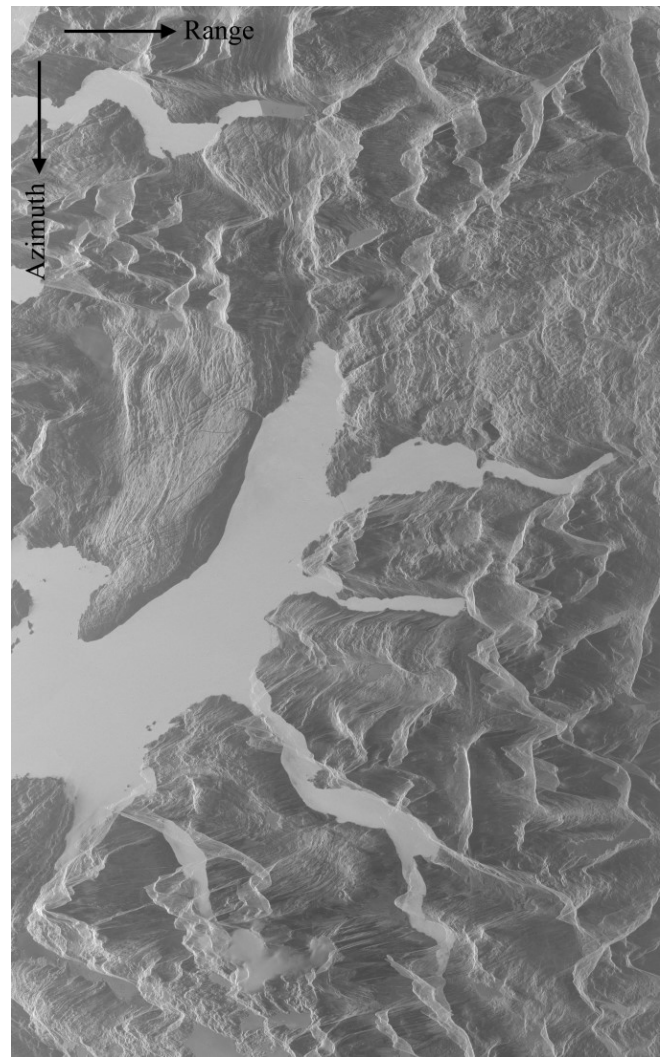

*Figure 1. An image acquired in Stripmap mode from ICEYE SAR satellite constellation over the Ofotfjord in Norway on February 24th, 2020 (size approximately 40 km x 70 km)*

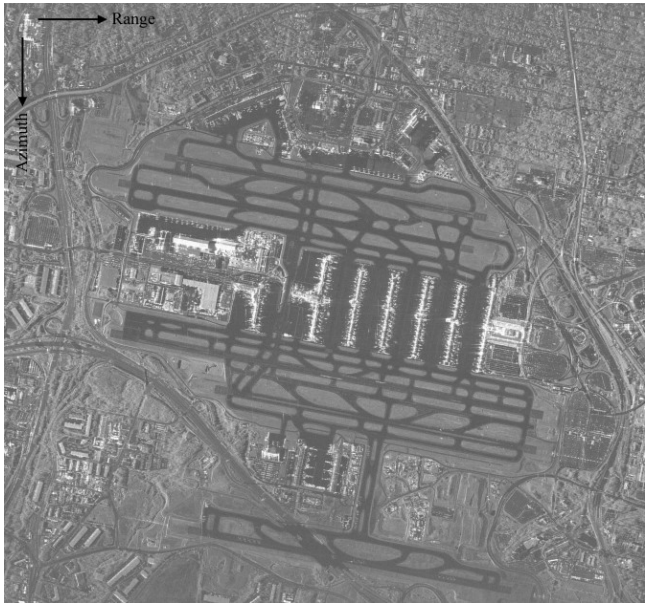

*Figure 2. A Spotlight mode ICEYE image with 25 cm resolution acquired over the airport of Atlanta in USA on March 24th, 2020 (size approximately 5 km x 5 km)*

*Table 2. Summary of ICEYE imaging modes*

| PARAMETER | STRIPMAP | SPOTLIGHT |
|---|---|---|
| Nominal swath width | 30 km | 5 km |
| Nominal product length | 50 km | 5 km |
| Look angles | 10-30 deg | 20-35 deg |
| NESZ | <- 17 db | <- 17 db |
| AASR & RASR | <-17 db | <-17 db |
| SLC, slant range resolution | 0.5 (300 MHz) 1.5 (100 MHz) | 0.5 (300 MHz) |
| SLC, azimuth resolution | 2.5 - 3.0 | 0.2-1.0 |
| SLC, range spacing | 0.4-1.3 | 0.4 |
| SLC, azimuth spacing | 1.4 - 1.7 | 0.35 - 0.7 |
| GRD, ground range resolution | 3.0 | 1.0 |
| GRD, azimuth resolution | 2.5-3.0 | 0.2-1.0 |
| GRD, range spacing | 2.5 | 0.5 |
| GRD, azimuth spacing | 2.5 | 0.5 |

Level-1 Single Look Complex products are basic single look products of the focused SAR signal. Scenes are stored in the satellite image acquisition geometry in the slant range by azimuth imaging plane in zero-Doppler SAR coordinates. The pixels are spaced equidistant in azimuth (according to the inverse of the pulse repetition frequency) and in slant range according to the range sampling frequency). Each image pixel is represented by a complex magnitude value (with inphase I and quadrature Q components) and therefore, contains both amplitude and phase information. Each image pixel is processed to zero Doppler coordinates in range direction, i.e. perpendicular to the flight track.

Level-1 Ground Range Detected (GRD) products represent focused SAR data that has been detected, multi-look processed and projected to the ground range using an Earth ellipsoid model. The image coordinates are oriented along the flight direction and along the ground range. The pixel spacing is equidistant in azimuth and in ground range. Ground range coordinates are the slant range coordinates projected onto the ellipsoid of the Earth. For the slant to ground range projection the WGS84 ellipsoid and a scene-averaged value of terrain height is used, annotated in the metadata. Pixel values represent detected magnitude. Phase information is lost. The resulting product has approximately square spatial resolution and square pixel spacing with reduced speckle due to the multi-look processing.

Slant range products are delivered in digital numbers and can be calibrated to radar brightness using annotated calibration factors. There is no correction for ground range projection or illumination effects. This reflects the fact that the calibrated brightness values relate to unit areas in slant range. The products are corrected for any measured or characterized gain variation including the ones of the instrument and resulting from projected antenna pattern. ICEYE SAR processor design and calibration and validation activities were performed using state-of-the-art approaches described for regular SAR missions [5, 6, 7, 8, 9].

The ICEYE SAR processor has a normalized signal gain for all imaging modes and a single absolute calibration constant is applicable to all products. The digital representation of the data is offered as int16.

The ICEYE SAR processor compensates the following effects:
- Range spread loss;
- Elevation antenna pattern;
- Azimuth antenna pattern in spotlight mode;
- Effects of different azimuth and range bandwidths;
- Sensor settings variations (receiver gain, transmit power, duty cycle)

### 3.2. Geometric Resolution

The best achievable slant range resolution of ICEYE-X2 is 0.5 meters based on the range bandwidth of 300 MHz. For SLC products no windowing is applied during the focusing to preserve the maximum spatial resolution. For detected products, the maximum resolution is somewhat reduced by weighting the range and azimuth spectrum with a tapering window to suppress the sidelobes of the Impulse Response Function (IRF) function to -17 dB or better. This side-lobe suppression is particularly important in imaging urban and industrial areas where the high spatial resolution of the system exposes high numbers of extremely strong scatters leading to high image contrasts. Additionally, the level of azimuth ambiguities is decreased. By enlarging the azimuth processing bandwidth in Stripmap mode, the resulting impulse response broadening can be compensated.

When transferring to ground range, the slant range resolution degrades by a factor of sin−1(incidence − angle). Both window-weighting and band-width-processing are performed simultaneously to optimize resolution, sidelobe suppression and ambiguity performance, with a goal of achieving approximately 3 m resolution in ground range. In azimuth, the theoretical resolution in Stripmap mode is half the antenna length. Due to finite sampling of the Doppler spectrum, aliasing effects are typically present. In the processor, the bandwidth reduction (during the SLC product formation) and spectral shaping (during the GRD product formation) is performed in order to reduce the ambiguities caused by aliasing and to improve the shape of the IRF. A constant resolution of 3 meters is a design goal for all ICEYE stripmap products. For ScanSAR, the need to ensure complete burst overlaps can slightly broaden the azimuth IRF. The processed Doppler bandwidth in Stripmap mode is currently 2700 - 3100 Hz.

### 3.3. IRF Quality Parameters

The quality of ICEYE SAR data has been assessed on a consistent SLC dataset via IRF (Impulse Response Function) measurements of Corner Reflectors (CR). The quality parameters of the IRF include the Peak Side Lobe Ratio (PSLR) and the Integrated Side Lobe Ratio (ISLR), that define how the focused energy is distributed spatially. It is important to keep in mind that SLC products are generated without using windowing functions during the data processing. The IRF quality parameters for each ICEYE satellite are gathered in Table 3, considering that the theoretical values for PSLR in range and azimuth are -13.2 dB and for ISLR in range and azimuth are -5.03 dB.

*Table 3. IRF Quality parameter statistics*

| Satellite | X2 | X4 | X5 |
|---|---|---|---|
| Rg PSLR [dB] | -13.71 ± 0.58 | -13.07 ± 0.94 | -13.44 ± 0.64 |
| Az PSLR [dB] | -13.5 ± 0.9 | -13.93 ± 1.06 | -13.79 ± 1.07 |
| Rg ISLR [dB] | -5.25 ± 0.39 | -5.04 ± 0.53 | -5.13 ± 0.47 |
| Az ISLR [dB] | -5.07 ± 0.38 | -5.11 ± 0.43 | 5.09 ± 0.38 |

### 3.3. Radiometric Performance

The absolute radiometric accuracy for the products (including all errors from calibration devices and processing) derived during the commissioning phases of both instruments and confirmed by calibration campaigns is less than 2 dB. Relative Radiometric Accuracy is the standard deviation of the radiometric error of known targets within one data take. Contributions come from the antenna pattern, unknown atmospheric absorption and chirp dispersion, the pointing knowledge of the antenna pattern and drifts of the instrument during operation. The relative radiometric accuracy estimate is better than 1 dB for ICEYE Stripmap mode data. Noise level (NESZ) is estimated to be better than -17 dB.

### 4. CONCLUSION

In this paper, we have presented the current status of the ICEYE SAR constellation and the description of the current system parameters, confirming the commercial maturity of "micro" sized instruments and platforms. The data quality has been ensured by a dedicated phase of calibration and validation performed on distributed targets (Amazon forest) and Corner Reflector sites. The constellation capabilities will be extended to the new satellites that will ensure that an operational availability with a higher revisit time will be achieved, allowing the near real time monitoring of phenomena with fast dynamics and enabling a large amount of applications such as SAR interferometry and Coherent Change Detection.